\documentstyle[12pt]{article}
\def\comma{\, , }
\def\period{\, .}
\def\nn{\nonumber}
\def\e{{\hat e}}
\def\ie{{$i.e.$}}

\def\a{{\alpha}}
\def\b{{\beta}}
\def\eps{{\epsilon}}
\def\n{{\eta}}

\def\s{{\sigma}}
\def\r{{\rho}}
\def\z{{\zeta}}
\def\x{{\xi}}
\def\d{{\delta}}
\def\t{{\theta}}
\def\l{{\lambda}}

\def\cd{{\cal D}}
\def\co{{\cal O}}
\def\cn{{\cal N}}
\def\cav{{\cal V}}
\def\cv{{\nu}}

\def\pr{{\partial}}
\def\prt{{\pr}_{\t}}
\def\tri{{\triangle}}

\def\S{{\Sigma}}
\def\G{{\Gamma }}
\def\sp{\vspace{.1in}}
\def\hs{\hspace{.25in}}
\def\fbar{\bar{f}}

\def\al{\alpha}

\def\Ttil{\tilde{T}}
\def\phitil{{\tilde{\phi}}}
\def\altilp{\tilde{\al}'}
\def\del{\partial}
\def\calL{{\cal L}}
\def\ep{\epsilon}
\def\calG{{\cal G}}
\def\matrixii#1#2#3#4            {  \left(\begin{array}{cc}#1&#2\\#3&#4
                                       \end{array}\right) }
\def\half       {  {1\over 2}  }

\def\calA{{\cal A}}
\def\calD{{\cal D}}
\newcommand{\be}{\begin{equation}} \newcommand{\ee}{\end{equation}}
\newcommand{\bea}{\begin{eqnarray}}\newcommand{\eea}
{\end{eqnarray}}
\renewcommand{\thefootnote}{\fnsymbol{footnote}}
\begin{document}
\begin{titlepage}
\begin{flushright}
{{UT-Komaba/98-17}\\
{hep-th/9807239}\\
{July, 1998}}
\end{flushright}
\begin{center}
\baselineskip= 26 truept
\vspace{.5in}
{\Large \bf  Interaction of D-string with F-string:\\
 A Path-Integral Formalism}

\vspace{.8in}
{\large Supriya Kar\footnote[2]{ supriya@hep1.c.u-tokyo.ac.jp;\ Address after
 October 1, Dept. of Theoretical Physics, Chalmers Institute of Technology
 \& Goteborg University, Sweden.}
 and Yoichi Kazama\footnote[3]{ kazama@hep1.c.u-tokyo.ac.jp}}
\end{center}
\begin{center}
\baselineskip= 18 truept

{\large \it Institute of Physics, University of Tokyo\\
 Komaba, Meguro-ku, Tokyo 153, Japan}

\end{center}

\vspace{0.4 in}
\baselineskip= 18 truept
\begin{center}
{\large\bf Abstract}
\end{center}

\vspace{.2in}

A path integral formalism is developed to study the interaction of an 
arbitrary curved  Dirichlet (D-) string with  elementary excitations 
of the fundumental (F-) string in bosonic string theory. 
Up to the next to leading order in the derivative expansion, 
we construct  the properly renormalized vertex operator, 
which generalizes the one previously obtained  for a D-particle
moving along a curved trajectory. 
Using this vertex, an attempt is further made to quantize the 
D-string coordinates and to compute the quantum 
amplitude for scattering between  elementary excitations 
of the D- and F-strings. By studying the dependence on the Liouville 
mode for the D-string, it is found that the vertex in our approximation 
consists of an infinite tower of local vertex operators which are 
conformally invariant on their respective mass-shell. 
This analysis indicates that, unlike the D-particle case, an off-shell 
extension of the interaction vertex would be necessary to compute 
the full amplitude and that the realization of symmetry
can be quite non-trivial when the dual extended objects are
simultaneously present. Possible future directions are suggested.
\thispagestyle{empty}
\end{titlepage}
\renewcommand{\thefootnote}{\alph{footnote}}
\baselineskip= 18 truept
\section{Introduction}

\hs
The idea of D-branes \cite{pol1,dai}
 has led to so many new results in the 
past few years that it is  now an indispensable part of our thinking 
in string-related areas \cite{pcj,bachasrev}. Nevertheless, when it 
comes to  quantum dynamics of D-branes, our knowledge is still far 
from complete: A considerable number of calculations have been performed 
to understand  the interaction between a D-brane and string states or between 
D-branes \cite{bachas}--\nocite{lifs}\nocite{barb}\nocite{douglas}
\nocite{klebthor}\nocite{gubseretal}\nocite{hashkleb}\nocite{garmyers}
\nocite{danfersun}\nocite{kabpou}\nocite{dkps}\nocite{calkleb}
\cite{li}  (and additional references in \cite{bachasrev}), 
but in most of these works D-branes are treated as infinitely 
heavy backgrounds. The non-linear dynamics of D-brane(s) in interaction 
with massless background fields is neatly coded in the Dirac-Born-Infeld 
action \cite{leigh,tseyt,callan} but its quantization is in general 
 difficult if not intractable. 
At low energies, this can be approximated by the celebrated super Yang-Mills
theory (SYM) on the worldvolume\cite{wit}, but again, in general, analysis of 
the quantum dynamics is not an easy task. \par
The exceptions are the cases of D-particles and D-instantons. For the former,
the SYM theory becomes a matrix quantum mechanics and 
moreover it was brilliantly reinterpreted as a promising candidate for
a microscopic description of M-theory in the light-cone frame\cite{bfss}. 
Its quantum properties have been under vigorous investigations and many 
results have already been obtained \cite{matrev}. 
Likewise, the SYM theory for D-instantons
was recognized as providing a non-perturbative definition of the type IIB 
superstring\cite{ikkt} and is being actively pursued. 
Progress on the quantum dynamics
of D-particles has also been made from the string theory point of view. 
One of us (Y.K.), in collaboration with S.Hirano, developed 
a path-integral formalism to quantize a D-particle in interaction with 
closed string states, thereby incorporating the recoil effects \cite{hk}. 
He further showed \cite{kaz}, with three complementary methods, 
how one can quantize the system of two D-particles 
 in string theory and obtained the fully 
quantum amplitude that generalizes the result obtained in \cite{bachas}. 

\sp
In this paper, we will focus on the dynamics of D-string 
in interaction with  F-string (\ie\  the usual string).
 The long range motivation behind this work is the desire to understand 
 the S-duality
of the IIB theory \cite{hultown}, one of the key symmetries of the string
 theory. 
Although a number of evidences exist, they are either classical or 
indirect. A more direct fully quantum mechanical demonstration would 
require that we be able to treat the D-string and the F-string 
 on equal footing. In particular, we should be able to quantize the 
excitations of D-string. This in turn requires, as a first step, a proper 
treatment of an arbitrary {\it curved} D-string interacting with F-string.

\sp
For this purpose,  we shall 
extend the path-integral formalism developed in \cite{hk} and derive 
the vertex operator describing the interaction of a curved D-string 
with excitations of F-string in bosonic string theory. The necessary 
 calculations are much more involved compared with the D-particle case, 
 but, employing  the derivative expansion which was successful in the 
 D-particle case,  we will be able to obtain a relatively compact result,
 very similar in form to the one for D-particle. 
 In this process, 
a short-distance divergence, which depends on the extrinsic curvature 
of the D-string, appears. It is gratifying that this divergence can be 
neatly absorbed by a renormalization of the D-string coordinates.
This is an important check of our formalism.

\sp 
Using the renormalized vertex operator so obtained, we will 
 make an attempt to compute the amplitude for the scattering 
 between elementary excitations of D- and F-strings. This  
 requires further path-integration over the D-string coordinates including 
 the additional vertex operators for the D-string excitations and 
 with the proper weight, \ie\  the exponential of the D-string action. 
In the case of  D-particles, the corresponding 
  program was successfully accomplished \cite{hk}, 
 despite the highly non-linear nature of the interaction vertex:
 The effect of the quantization could be summarized in 
a simple rule that the (proper-)time 
derivative $\dot{f}^\mu$ of the D-particle coordinate 
be replaced by the average ${1\over 2}(p^\mu+{p'}^\mu)$
of the incoming and the outgoing momenta of the D-particle, which 
 had been first conjectured in \cite{ishibashi}. 
 In the present case of D-string, the situation turned out to be much more
 involved because of the stringent requirement of the conformal invariance 
 which is absent in the D-particle case. 
The analysis of the Liouville mode of the D-string in the conformal gauge 
 indicates that our approximation only picks up the on-shell intermediate 
 states and the full conformal invariance is not achieved unless an 
 appropriate off-shell extension of the vertex will be made. 
 We may draw from this  an important lesson that the realization of 
 symmetry, such as the conformal invariance in the present case, 
can be  highly non-trivial when  the dual extended objects are  simultaneously
 present. 

\sp
The organization of the rest of the article is as follows:  In Sec.2, we 
describe the general setup of the problem, including the explanation 
 of the geodesic normal coordinate expansion, the orthonormal moving 
 frame to be used, and the constraints at the boundary of the 
 open string world-sheet. The actual computation begins in Sec.3, where 
 the path-integral over the open string coordinates is performed. The 
 remaining integrations over the fields at the boundary will be 
 treated in Sec.4. In Sec.5, the renormalization of the vertex operator 
 is conducted and the its $SL(2,R)$ transformation property is explained. 
 We then describe, in Sec.6, an attempt to compute the 
 amplitude for the scattering of excitations of D- and F- strings. After 
 a brief r{\'e}sum{\'e} of the quantization procedure, we examine 
 the question of conformal invariance of the vertex operator, analyze the 
 nature of the problem, and suggest future directions.
\section{Path integral formalism}
\subsection{The general setup}

\hs
We begin with the characterization of an arbitrary curved D-string. 
Let $f^\mu(t,\s)$ be the coordinates of a curved D-string embedded 
in a flat space-time, with its world-sheet  described by $(t,\s)$. In 
this article, we will take the topology of the worldsheet of the open string
attched to the  D-string to be a disk $\S$ and parametrize its boundary 
$\pr\S$ by the polar angle $\t \,, ( 0\leq \t \leq 2\pi )$. Since the ends 
of the open string may lie anywhere on the world-sheet of the D-string, 
the Lorentz covariant condition at the boundary which characterizes
the D-string is given by \cite{leigh}

\be 
X^{\mu}({\t})\ = \ f^{\mu}\Big ( t({\t}) , {\s}({\t})\Big ) \ ,\label{dsdef}
\ee
where $X^{\mu}$ denote the open string coordinates,  and 
$t(\t )$ and $\s (\t )$ are arbitrary functions describing where on the 
world-sheet the ends of the open string land. \par
The object of our interest is the vertex operator (\ie\ a functional 
 of the D-string coordinates $f^\mu(\t,\s)$)  for a D-string interacting
with the states of closed F-string. In the path integral formalism, 
it is obtained by integration of an appropriate weight over $X^\mu(z)$, 
$t(\t)$ and $\s (\t)$ with the constraint (\ref{dsdef}) imposed,
 together with the insertions of the vertex operators for the closed 
F-string states. Also, as usual, we will include 
in our formalism the coupling of the ends of the open string to the 
abelian gauge field $A_\mu(X)$. Then the precise form of the vertex 
operator to be studied is 
\bea
{\cal V} \Big ( {f^{\mu}} , \{{k_i}\} \Big )&=&
{1\over{g_s}} \int \cd X^{\mu}(z,\bar z )
\ \cd t({\t}) \ {\cd \s}({\t})\ {\d } \Big (\ X^{\mu}(\t) - f^{\mu} 
(t(\t) , {\s}(\t))\ \Big ) \nonumber \\
&& {} \qquad\qquad\qquad\qquad
\cdot \exp \Big (-S[X,A]\ \Big )\ \prod_i {g}_s \ V_i ({k_i}) \ ,
\label{defvertex}
\eea
where $g_s$ is the string coupling constant and $V_i ({k_i})$ are the vertex
operators for closed string states carrying momenta $k_i$.
For definiteness, we will consider the tachyon emission vertices
$V_i(k_i) = \int d^2 z_i \exp\Big ( i k_i\cdot X(z_i) \Big )$. 
$S[X,A]$ is the open string action coupled to the gauge field and is 
of the familiar form\cite{tseyt,callan}
\be
S[X,A]\ = \ {1\over{4\pi\a'}} {\int}_{\S} d^2z \  {\pr_{\bar a}} X^{\mu}
{\pr_{\bar a}} X_{\mu} \ + \ i {\int}_{\pr {\S}} d\t \ A_{\mu} (X) 
\ {\prt} X^{\mu} \ ,\label{action}
\ee
where $\bar a = 0 , 1$ labels the coordinates on the F-string world-sheet. 
\subsection{Geodesic normal coordinate expansion}

\hs
Let us first consider the integrals over $t(\t)$ and $\s(\t)$, 
describing the fluctuations of the ends of the F-string. 
Although these fluctuations play an important role in producing the
  effective action of the D-string, 
they cannot be dealt with exactly. Hence, as in the case of a 
D-particle \cite{hk}, 
 we will employ the geodesic normal coordinate expansion \cite{alvarez}
 and organize 
 their effects  order by order in the derivatives of the 
D-string world-sheet coordinates $f^\mu(t,\s )$. 
Clearly such an expansion preserves the general coordinate invariance 
on the world-sheet of the D-string.
\par
Adapting the method of \cite{alvarez} to our case, the expansion of 
$f^{\mu}( t(\t),\s (\t ))$ around 
$f^{\mu}(t,{{\s }})$ is worked out as 
\bea
f^{\mu}\pmatrix {{t(\t),\s(\t) } \cr }&=& f^{\mu} ( {t,\s} ) \
+ \ {\pr }_a f^{\mu} (t,\s ) {\z }^a ({\t }) 
\ + \ {1\over2} K^{\mu}_{ab}
{\z }^a({\t }) {\z }^b({\t }) \nonumber \\
&&{}\qquad\qquad\qquad
+ \ {1\over{3!}} K^{\mu}_{abc} {\z }^a({\t }) 
{\z }^b(\t ) {\z }^c(\t ) \ + \ {\co}({\z }^4 ) \ ,
\eea
where ${\z }^a({\t })$ are the normal coordinates with  $a=(t,\s )$
 and $K^{\mu}_{ab}(t,\s )$ and $K^{\mu}_{abc}(t, \s )$ are the extrinsic
curvatures of the D-string sub-manifold.
 They can be expressed as 

\bea
K^{\mu}_{ab} \ &=& \ P^{\mu\nu}\ \pr_a\pr_b f_{\nu}\ , \\
K^{\mu}_{abc} \ &= &\ \pr_a\pr_b\pr_c f^{\mu} - \pr_a {\G^{\mu}}_{bc}
+2 {\G^d}_{ab} {\G^{\mu}}_{dc} \ ,
\eea
where $P^{\mu\nu}$ is the projection operator normal to the world-sheet and
$\G^a{}_{bc}$ is the Christoffel connection.
More explicitly, $P^{\mu\nu}$ is given in terms of the projector $h^{\mu\nu}$
tangential to  the world-sheet as 

\be
\n^{\mu\nu} \ = \ h^{\mu\nu}\ + \ P^{\mu\nu} \ ,
\ee
where $h^{\mu\nu}$ is constructed out of the inverse $h^{ab}$ of 
the metric $h_{ab}$ induced on the world-sheet of the D-string:
\bea
h^{\mu\nu} \ &=& \ {\pr }_a f^{\mu}h^{ab}{\pr }_b f^{\nu} \ ,\\
h_{ab} \ &=& \ {\pr }_a f^{\mu}{\pr }_b f_{\mu} \ .
\eea
The intuitive picture of this expansion is that  to the leading order 
 the entire boundary of the disk is
attached to a point $(t,\sigma)$ on the D-string world-sheet 
and the effects of the 
fluctuations from this configuration is  taken into account 
 by the integration over the normal coordinates $\z^a(\t )$.
\par
Now to facilitate the computation, we split the string coordinate 
$X^{\mu}(z, \bar z )$ into a constant 
and  non-constant modes: 

\be
X^{\mu}(z, \bar z ) \ = \ x^{\mu} \ + \ {\x}^{\mu} (z, \bar z ) \ .
\ee
Then the $\d$-function expressing the constraint (\ref{dsdef}) 
 on the boundary decomposes into two parts:
\bea
\d \Big ( X^{\mu}(\t) - f^{\mu}(t(\t),\s(\t) ) \Big )
&=& \d \Big ( \ x^{\mu} - f^{\mu} (t , {\s}) \Big )\nonumber \\
&& {}\qquad 
\cdot \d \Big ( {\x }^{\mu} (\t ) - \pr_a f^{\mu}\Big ( t, \s \Big )
{\z }^a (\t ) - \dots \ \Big ) \ .\label{bconst}
\eea
The first $\delta$-function makes the integration over $x^\mu$ trivial 
and produces, 
from the product of tachyon vertices, 
a factor 

\be
V_0 = \exp \Big ( i \ k^{\mu} f_{\mu}(t , \s )\ \Big )\ ,
\ee 
where $k^{\mu}\ \equiv \ \sum_i k_i^{\mu}$  is the total momentum of 
the tachyons.  The effect of the second $\delta$-function will be discussed 
 later in Sec.3.1. 

\subsection{Orthonormal frame}

\hs
Just as in the D-particle case\cite{hk}, integration over the 
non-constant mode $\xi^\mu(z,\bar z )$ will be simplified by the use of 
an appropriate orthonormal moving frame. Since, to the zero-th
order in the normal coordinate expansion, the entire boundary of the 
disk is mapped on to a point $(t,\s )$ on the D-string world-sheet, 
 the most natural choice is  a frame where two of the 
basis vectors,\ $\e^{\mu}_a, \, a=0,1$,  lie on the tangential plane 
at that point, with  the rest, $\e^{\mu}_\alpha, \, \alpha = 2,3,\ldots,
25$,  being orthogonal to them. In this way, the boundary conditions for the 
open string will be simply Neumann along $\e^{\mu}_a$ and Dirichlet 
along $\e^{\mu}_\alpha$, to this order of approximation.

\sp
Specifically,  we construct the orthonormal vectors ${\e^{\mu}_A}$ 
for $A = (a, \a )$ in the following way:

\bea
{\e^{\mu}}_0 &=& {{{\dot f}^{\mu}}\over{\sqrt{-h_{00}}}}\comma \qquad 
{\e^{\mu}}_1  {\sqrt{{h_{00}}\over{h}}} \ 
\Big ({f'}^{\mu} \ + \ \e^{\mu}_0
\e^{\nu}_0 \ {f'}_{\nu} \Big ) \ , \nonumber \\
\e^{\mu}_A \e_{\mu B} &=& \n_{AB} \ , \nonumber\\
\n^{\mu\nu} &=& {\e^{\mu}}_A {\e^{\nu}}_B \ \n^{AB} \ , \\ 
h^{\mu\nu}  &=&  \sum_a {\e^{\mu}}_a {\e^{\nu}}_a  
=  - {\e^{\mu}}_0 {\e^{\nu}}_0 \ + \ {\e^{\mu}}_1 
{\e^{\nu}}_1 \ , \nonumber \\
P^{\mu\nu} &=& \sum_{\a} {\e^{\mu}}_{\a} {\e^{\nu}}_{\a} \ .\nn
\eea
Here, $h$ stands for $\det h_{ab}$, 
a dot and a prime on $f^{\mu}$ correspond to time $(t)$ and spatial $(\s )$
derivatives respectively, and we have displayed as well the expressions for 
the projectors in terms of the basis vectors. We will actually use 
more compact notations given by 

\bea
\e^{\mu}_a &\equiv & \cn_a\ \pr_a f^{\mu}(t, \s ) \ , \nonumber \\
\e^{\mu}_A & \equiv & \cn_A \ e^{\mu}_A \ ,
\eea
where $a= (0,1) $, $\cn_0 = {1\over{\sqrt{-h_{00}}}} $,
$\cn_1 = {\sqrt{{h_{00}}\over{h}}}$ and 
$\cn_A \equiv ( \cn_a , 1,1,\dots  1 )$.
(Apart from being orthogonal to $\e^\mu_a$, explicit forms for $\e^\mu_\alpha$
will not be needed.)
Now in this frame, the non-constant mode
$\x^{\mu}(z, \bar z )$ can be expanded as 

\be
\x^{\mu}(z, \bar z )\ = \ \sum_A \e^{\mu}_A \ \r^A (z, \bar z ) \ ,
\ee
where $\r_a(z,\bar z ) $ and $\r_{\a}(z, \bar z ) $ correspond to the 
fluctuations in the tangential
and the transverse directions respectively. 

\subsection{\lq\lq Constraints" and \lq\lq conditions" at the boundary}

\hs
Before starting the path integration, we must clarify the conditions 
imposed at the boundary and how we treat them.
There are two types of such conditions governing the fluctuations of 
the string coordinates $\r_A(\t)$ and the geodesic coordinates $\zeta^a(\t)$.

\sp
The first set of conditions, which we call the \lq\lq boundary constraints",
come from the $\delta$-function in Eq.(\ref{bconst}). Expressed as relations 
between $\r_A(\t)$ and $\zeta^a(\t)$, they take the form 

\bea
&&\r_a(\t ) \ = \ 
\ \cn_a^{-1} \ \n_{ab} \Big ( \ \z^b(\t ) \ - \ {1\over{3!}} \Big [
K^{\l}_{lm} K_{\l np} \ h^{bp} \ + \ \G^b_{lp} \G^p_{mn} \
+ \ \pr_l\pr_m f^{\l} \pr_q f_{\l} \ \pr_n h^{bq} \nonumber \\
&& \qquad 
+ \ 2 \pr_l \pr_m 
f^{\l} \pr_q f_{\l} \ \G^q_{np} h^{bp} 
- \ \pr_r f^{\l} \pr_q f_{\l}
\ \G^q_{np} \G^r_{lm} h^{bp} \ \Big ] \z^l(\t ) \z^m(\t ) \z^n(\t )
\Big ) \ + \ \co ( \z^4 )\nn\\
\eea
and
\bea
\r_{\a}(\t ) &=& \e^{\l}_{\a} \Big ( \ {1\over2} \pr_a \pr_b f_{\l} 
\ \z^a(\t )\z^b(\t ) 
+ \ {1\over{3!}} \Big [ \pr_a \pr_b \pr_c f_{\l} \
-\ 3 \G^d_{bc}\ \pr_d \pr_a f_{\l} \Big ] \nonumber \\ 
&& {\hspace{1in}}  \cdot \z^a(\t ) \z^b(\t ) \z^c(\t ) 
\ \Big ) 
+ \ \co (\z^4 ) \ .
\eea
These are much more complicated than their counterparts in the 
D-particle case\cite{hk} due to the nature of the  general two dimensional 
induced metric on the D-string sub-manifold. We will regard them as 
constraining the integrations over $\zeta^a(\t)$, which will be performed 
after integrating over $\r_A(\t)$.

\sp
The second type of conditions  
are the usual boundary conditions for $\r_A(\t)$ that arise from 
the consistency of the variation of the action. To find them 
we need to expand the gauge field $A_\mu(X)$ in Eq.(\ref{action}) 
around the constant mode $x^\mu$. To avoid unnecessary complications, 
we will deal  with the case of constant field strength \cite{callan},
 for which the boundary interaction becomes quadratic in $\xi^\mu$:

\be
i \ \int_{\pr\S} d\t \ A_{\mu} \Big ( x + \x \Big ) \ \pr_{\t} X^{\mu}
\ = \ {{i}\over{2}}\ F_{\mu\nu} \ 
{\int}_{\pr\S} d\t \ \x^{\mu} \pr_{\t} \x^{\nu} \ .
\ee
Then the action (\ref{action}) in the orthonormal frame becomes 

\bea
S[X,A] &=& {1\over{4\pi\a'}} \Big [ - {\int}_{\S} d^2z \ \r_A \pr^2 \r^A
\nonumber \\
&& {} \qquad\quad + \ {\int}_{\pr\S} d\t \ \Big ( \ \r_A \pr_n \r^A \ 
+ \ i\ \cn_A \cn_B {\bar F}_{AB} \ \r^A
\pr_{\t} \r^B \ \Big )\ \Big ] \ ,\label{raction}
\eea
where we introduced the notations ${\bar F}_{AB} \equiv 2\pi\a' \, F_{AB}$,
$\pr_A \equiv e^{\nu}_A \pr_{\nu}$ and  
$A_B \equiv e^{\mu}_B A_{\mu}$. From this action, it is straightforward 
to obtain the following set of consistent boundary conditions:
\bea 
&&\pr_n \r_a (\t ) \ + \ i \ \cn_a \cn_b \
{\bar F}_{ab} \ \pr_{\t}\r^b(\t ) \ = \ 0\comma 
\nonumber \\
&& \r_{\a}(\t ) \ = \ 0 \ .\label{bcond}
\eea 
Note that, as expected, the D-string only sees $\bar{F}_{ab}$, the 
components of the gauge field along its world-sheet. 

\section{Path integral over the string coordinates $\r^A(z, \bar z )$}

\hs
We are now ready to perform the integration over the non-zero modes 
$\r_A(z,\bar{z})$, which exist on the boundary as well as in the bulk. 
Rather than treating the boundary components separately, we shall 
regard them as bulk quantities with appropriate $\delta$-functions, so
that the integration can be performed in a unified manner.

\subsection{Reformulation of the boundary interactions}

\hs
 Let us describe this 
procedure more explicitly. 
First consider the non-zero mode part of the 
$\d$-function constraints  (\ref{bconst}). It can be rewritten 
as an integral over a Lagrange multiplier $\cv^{\mu}(\t )$ in the 
form 

\bea
&&\d \Big  ( \x^{\mu}(\t )\ - \ \pr_a f^{\mu} \ \z^a(\t ) \
- \ {1\over2} K^{\mu}_{ab}\ \z^a(\t ) \z^b(\t ) \ - \ {1\over{3!}} 
K^{\mu}_{abc} \ \z^a(\t )\z^b(\t ) \z^c(\t ) \ \dots  \ \Big ) \nonumber \\
&& \qquad = \ \int \cd \cv^{\mu}(\t )\ \exp \left (\ i \int d\t \
\cv_{\mu}(\t )\ \x^{\mu}(\t )\ \right ) \
\cdot \exp 
\Big ( -i \int d\t \ \cv_{\mu}(\t )\ \Big [ \ \pr_a f^{\mu} \z^a(\t ) 
\nonumber \\
&& \qquad\qquad\qquad + \ {1\over2} {K^{\mu}}_{ab} \z^a(\t ) \z^b(\t ) 
\ + \ {1\over{3!}} {K^{\mu}}_{abc} \z^a(\t ) \z^b(\t ) \z^c(\t ) \ + \dots \ 
\Big ] \ \Big ) \ ,
\eea
where $\cv^A(\t ) \equiv \cv_{\mu} \e^{\mu A}$ are the components 
in the orthonormal frame. Together with  the use of the 
boundary condition (\ref{bcond}) the above expression can be written 
explicitly as

\bea
&& \ \int \ \cd\cv^a(\t ) \ \cd\cv^{\a}(\t )
\ \exp \Big (\ i \int d\t \ \cv^a(\t )\
\r_a (\t ) \ \Big )
\cdot \exp 
\Big (\ - i \int d\t \ \cn_a^{-1}  \ \n_{ab} \ \cv_b(\t )\nonumber \\
&& \hspace{.6in} 
\cdot \Big [\ \z^b(\t ) \ - \ {1\over{3!}} \Big ( \ {K^{\l}}_{lm} {K_{\l}}_{np} 
\ h^{bp} \ 
+ \ \G^b_{lp}\ \G^p_{mn} 
\ + \ \pr_l \pr_m f^{\l} \pr_q f_{\l} \ \pr_n h^{bq} 
\nonumber \\
&& \hspace{.8in}
+ \ 2 \ \pr_l \pr_m f^{\l} \pr_q f_{\l}\  \G^q_{np} h^{bp} \
- \ \pr_r f^{\l} \pr_q f_{\l}\ \G^q_{np} \G^r_{lm}\ h^{bp} \Big ) \z^l(\t )
\z^m(\t ) \z^n(\t ) \Big ] \Big ) \nonumber \\
&& \hspace{.6in}
\cdot \exp \Big ( -i \int d\t \ \cv^{\a}(\t )\ \e^{\l}_{\a} \Big [\ {1\over2}
\ \pr_a \pr_b f_{\l}\ \z^a(\t ) \z^b(\t ) \nonumber \\
&& \hspace{.8in} - \ {1\over{3!}}\ \Big ( \pr_a \pr_b
\pr_c f_{\l} 
-\ 3\ \G^d_{bc}\ \pr_d \pr_a f_{\l} \ \Big ) \ \z^a(\t ) \z^b(\t ) \z^c(\t )
\ \Big ] \ + \ \co (\z^4 ) \  \Big ) \ .\label{deltafn}
\eea
The boundary value of $\r_A(z,\bar{z})$ appears only in the first 
exponent, which can be reexpressed as a bulk integral
$ i \int_{\S} d^2z \ \d_{Aa} \ \d \Big ( |z| - 1 \Big ) \ \cv^A(z,\bar z)
\ \r_A(z,\bar z' ) \ .$

\sp
Next consider the boundary interaction with the gauge field 
appearing in (\ref{raction}). Since the  
transverse coordinates $\r^{\a}(\t )$ drop out due to  
the boundary conditions (\ref{bcond}), we can write it in the bulk 
form as 

\bea
&& i \ \cn_A \cn_B \ {\bar F}_{AB}(x) \ {\int }_{\pr\S } 
d\t \ \r^A(\t ) \pr_{\t } \r^B(\t )\nonumber \\
&&\hspace{1in} = \ - \ \cn_a \cn_b \ {\bar F}_{ab}(x) \
{\int}_{\S} d^2 z\ 
\d \Big(|z| - 1 \Big ) \r^a(z,\bar z ) \
\pr_z\r^b (z, \bar z ) \ .\label{gint}
\eea
Now from Eqs.(\ref{raction}),(\ref{deltafn}) and (\ref{gint}), the 
path integral over $\r_A(z, \bar z )$ with tachyon vertex insertions
can be written as 

\bea
I_{\r}&=& \int \cd \r_A \ \exp \Big ( - {1\over{4\pi\a'}}
{\int }_{\S }
d^2 z \ \Big [\pr_{\bar a} \r^A \pr^{\bar a} \r_A
\ - \ \cn_A \cn_B  {\bar F}_{AB}(x)
\ \d \Big (|z|-1\Big ) \ \r^A \pr_z \r^B \Big ] \ \Big )
\nonumber \\
&& \hspace{2in}
\cdot \exp \Big ( \ i {\int }_{\S} d^2z \ J_A \ \r^A \ \Big ) \ ,
\eea
where
\be
J_A \ = \ \sum_i \ k_{iA} \ \d^{(2)}(z - z_i ) \ + \ \d_{Aa} 
\ \d \Big ( |z| - 1\Big )\ \cv_a(\t ) \ 
\ee
and $k_{iA} \ = \ k_{i\mu } \hat{e}^{\mu}_A $ are the momenta of the 
tachyons expressed in the orthonormal frame.
\subsection{Evaluation of $\r^A$-integrals}

\hs
Clearly the integrals over the transverse coordinates $\r^\alpha$, 
satisfying the Dirichlet boundary condition, 
can be performed rather trivially with the use of the well-known 
Dirichlet function $D(z,z')$ on the unit disk. We write the result as 

\be
I_\r^t =  \exp \left (\ {{\a'}\over2} {\int} \ d^2z\ d^2z' 
\ J_{\a}(z) G_{\a\a}(z,z') J_{\a}(z') \ \d_{\a\a} \ \right ) \ .
\ee
where 

\be
G_{\a\a}(z,z') \ = \ D(z,z') 
= \ \ln |z - z'| \ - \ \ln |1 - z {\bar {z'}}| \ .
\ee
The remaining integral over the longitudinal components $\r^a(z,z')$ is of the
form

\bea
I^l_{\r} \ &=& \ \int \cd \r_a \ \exp \left ( \ {1\over{4\pi\a'}}
 \int \ d^2z \ 
\r^a(z) \ \tri_{ab}\ \r^{b}(z)
\ + \ i \int \ d^2z \ J_a(z)\ \r^a(z) \ \right ) \ , \label{lro}
\eea
where $\tri_{ab}$ is a non-trivial operator given by

\be
\tri_{ab} \ = \ \n_{ab} \ \pr^2 \ + \ \cn_a \cn_b \ {\bar F}_{ab} 
\ \d\Big ( |z| - 1 \Big ) \ \pr_z \ .\label{delab}
\ee

The path integral (\ref{lro}) is easily done, with the result 

\bea
I^I_\rho &=& \left( -\det \tri_{ab} \right)^{-1/2}\, 
 \int \, \exp \left( {\a' \over 2} \int \, d^2z \, d^2 z'\, 
 J_a(z) \, G_{ab}(z, z' ) \, J_b(z' ) \, \eta_{ab} \right) \, , 
\label{intj}
\eea
where ${\bar F}_{ab} = \eps_{ab}{\bar f}$ with $\eps_{01} = 1 = -\eps_{10}.$
The propagator matrix $G_{ab}(z,z' )$ in Eq.(\ref{intj}) 
for $ a =(0,1)$ is the 
Neumann function on the unit disk and in the bulk it satisfies

\bea
\tri_{ab} G_{ab}(z , z' ) &=& \pr^2 G_{ab}(z,z' ) 
= 2\pi \ \n_{ab}\ \d^{(2)}(z,z' ) \ .\label{lbulk}
\eea
Also, due to the boundary condition (\ref{bcond}),
 $G_{ab}$ satisfies
\be
\pr_n G_{ab}(z, z' ) \ + \ i \ \cn_a \cn_b \
{\bar F}_{ab} \ \pr_{\t} G_{ab}(z,z' ) 
\ = \ 0 \ , \label{lbound}
\ee
on the boundary $\pr\S$. Explicitly, they are given in matrix form as

\bea
G_{ab}(z, z' )&=& \d_{ab} \ \ln |z - z'| + {1\over2} \left( 
{{{\sqrt{-h}} - {\bar F}}\over{{\sqrt{-h}} + {\bar F}}}
\right) _{ab} \ \ln \left ( 1-{1\over{z {\bar z'}}} \right ) \nonumber \\
&&\hspace{2in}
+ \ {1\over2} \left( {{{\sqrt{-h}} + {\bar F}}\over{{\sqrt{-h}} - {\bar F}}}
\right) _{ab} \ \ln \left ( 1-{1\over{z' {\bar z}}} \right ) \ .
\eea
It can be checked that 
$G_{00}(z,z' ) = G_{11}(z,z' ) \equiv G(z,z')$ and can be
expressed as

\be
G(z,z')  = \ln |z-z' | \ 
+ \ \left ( {{h+{\bar f}^2}\over{h-{\bar f}^2}}
\right ) \ln \left | 1-{1\over{z{\bar z}'}} \right | \ . \label{diagG}
\ee
For $ F_{ab} =0 $, the off-diagonal part of the
Neumann function vanishes and  the diagonal element (\ref{diagG}) 
corresponds exactly to that of a D-particle \cite{hk} in the mutually 
orthogonal directions.

\sp
On the  boundary $\pr\S,$   the diagonal part of the 
Neumann function $G(\t,\t')$ diverges  
 as $\t' \rightarrow \t$ and needs to be regularized. We adopt the 
 usual method 
\cite{tseyt} with a cut off $\epsilon$ and write 

\be
G(\t , \t' ) \ = \ -2  h  \Big (h-{\bar f}^2 \Big )^{-1} 
\ \sum_{n=1}^{\infty}
\ {{e^{-\eps n}}\over n} \cos\ n(\t -\t' ) \ .\label{Gtheta}
\ee
Then its inverse $G^{-1}(\t,\t')$  satisfies

\bea
&&{1\over2} \int \ d\t \ d\t' 
\ G^{-1}( \t , \t' )\ G( \t , \t' ) \ = \ - 1 
\nonumber \\
 && {1\over{2\pi^2}} \ 
G( \t , \t' ) \ \pr_{\t} \pr_{\t'} \ G( \t , \t' )\ 
\ = \ \tilde\d (\t - \t' ) \ .
\eea
where $\tilde\d (\t - \t' )=\delta(\t -\t') -(1/2\pi)$. 
With this regularization, $G(\t,\t)$ is given by 

\be
G( \t , \t ) \ = \ 2 \ 
\ h \ \Big ( h - {\bar f}^2 \Big )^{-1} \ln \eps \ .
\ee
This divergence will be seen to be absorbed by the renormalization 
 of the D-string coordinates. 

\sp
Finally,  we need to evaluate the Jacobian factor in 
 Eq.(\ref{intj}). Using the Fourier mode expansion on the 
 boundary circle and the usual $\zeta$-function 
 regularization, we find 

\bea
\Big ( - \det \tri_{ab}\ \Big )^{-{1\over2}} &=&  \prod_{n=1}^{\infty} \
\left ( {{h + {\bar f}^2}\over{h}} \right )^{-1}
=
\exp \left [ -\z(0) \ \ln 
\left ( {{h + {\bar f}^2}\over{h}}\right ) \ \right ] \ 
\nonumber \\
&=& \left ( \ {{h + {\bar f}^2}\over{h}} \ \right )^{1\over2} \ .
\eea

\sp
This completes the path integral over the string coordinates  $\r_A(z,z' )$.
 Putting everything together, the result takes the form 
\bea
I_{\r} &=& {\sqrt{{h + {\bar f}^2}\over{h}}} \ 
\cdot \exp \left ( \ {{\a'}\over2} \ 
\sum_{ij} k_{ia} 
k_{jb} \ G_{ab}(z_i , z_j ) \ \n_{ab} \ \right ) \nonumber \\
&& \qquad\qquad
\cdot \exp \left ( \ {\a'} \int d\t \ \sum_i k_{ia} \ G_{ab}(z_i , \t )
\cv_b(\t ) \ \n_{ab} \right ) \nonumber \\
&& \qquad\qquad
\cdot \exp \left ( \ {{\a'}\over2} \int d\t \ d\t' \ \cv_a(\t ) \
G_{ab}(\t , \t' ) \ \cv_b (\t' ) \ \n_{ab} \ \right ) \nonumber \\
&& 
\qquad\qquad
\cdot \exp 
\left ( \ {{\a'}\over2} \sum_{ij} k_{i\a} \ k_{j\b} \ G_{\a\b} (z_i , z_j )
\ \d_{\a\b } \ \right ) \ .\label{bulkint}
\eea
\section{Path integral over the boundary fields}

\hs
What remains to be performed is the integral over the 
 boundary fields, namely over the 
 Lagrange multiplier 
fields $\cv_a (\t ) $ and the normal coordinates $\z_a(\t )$. 

\subsection{Integration over the Lagrange multiplier $\cv_a(\t ) $}

\hs
Assembling the relevant terms from
Eqs.(\ref{deltafn}) and (\ref{bulkint}), the path integral over the 
Lagrange multipliers $\cv_a(\t ) $ takes  the form 

\bea
I_{\cv} &=& \int   
\ \cd \cv_a(\t ) \ 
\cdot \exp \left ( \ {{\a'}\over2} \int d\t \ d\t' \ \cv_a(\t )\ 
G_{ab}(\t , \t' ) \ \cv_b(\t' ) \ \n_{ab} \ \right )
\nonumber \\
&& \qquad\qquad\qquad\qquad 
\cdot \exp \left ( \ i\int \ d\t \ j_a(\t )
\ \cv_b(\t ) \ \n_{ab} \  \right ) \ ,\label{intnu}
\eea
where the source term is given by 
\bea
 {{j_a(\t )}\over{\sqrt{\a'}}} \ 
&=& \ \n_{ab} 
\ \Big ( -i{\sqrt{\a'}}\ \sum_i k_{ia} \ G_{ab}(z_i , \t )
\ + \ {{\cn_a^{-1}}\over{\sqrt{\a'}}} \ 
\Big [ \z^b(\t ) \nonumber \\
&& - \ {1\over{3!}} \Big ( K^{\l}_{lm} K_{\l np} \
h^{bp} \
+ \ \G^b_{lp} \ \G^p_{mn} \
+ \ \pr_l \pr_m f^{\l}\pr_q f_{\l} \ \pr_n h^{bq} \
+ \ 2 \pr_l \pr_m f^{\l} \pr_q f_{\l} \ \G^q_{np} h^{bp} \nonumber\\
&& - \ \pr_r f^{\l}
\pr_q f_{\l} \ \G^q_{np} \G^r_{lm} \ h^{ap} \ \Big ) \z^l(\t ) \z^m(\t ) 
\z^n(\t ) \ \Big ] \ \Big ) \ 
+ \ \co (\z^4 ) \ .
\eea
The integration is straightforward,  with the result 
\be
I_{\cv} \ \equiv \ 
\exp \left ( \ {1\over2} \int \ d\t \ d\t' 
\ {\tilde j}_a(\t ) \ G^{-1}_{ab}(\t ,\t' )
\ {\tilde j}_b(\t' ) \ \n_{ab} \right ) \ ,
\ee
where ${\tilde j}_a(\t ) \ \equiv \ {j_a(\t )}/{\sqrt{\a'}} $.
\subsection{Integration over the  normal coordinates $\z_a(\t )$}

\hs
Now we come to the final stage of the calculation, namely to the integration 
 over the normal coordinates $\z_a(\t )$. To simplify the calculation, 
 it is convenient to make a rescaling 
\be
\z^a(\t ) \ = \ {\sqrt{\a'}} \ \cn_a \ {\bar \z}^a(\t ).
\ee
This induces a change in the functional measure, which can be 
 computed just like for $(-{\rm det}\, \Delta_{ab})^{-1/2}$, with the result 

\be
\cd \z^a(\t ) \ = \ {1\over{\a'}}\ \ {\sqrt{-h}} 
\ \cd {\bar \z}^a(\t )\ .
\ee
Then, assembling all the relevant terms from (\ref{intnu}), (\ref{bulkint})
 and 
(\ref{deltafn}), the integral over ${\bar \zeta}(\t)$,  to 
 order $\co ({\bar \zeta}^4)$, becomes 
\bea
I_{\bar\z} &\equiv & \int \
\cd {\bar\z}_a \ \exp \left ( \ {1\over2} \int d\t \ d\t' \
{\bar\z}^a(\t ) \ G_{ab}^{-1}(\t ,\t' ) \ {\bar\z}^b(\t' ) \ \n_{ab} \ \right )
\nonumber \\
&& \cdot \exp \left ( \ -{{i\a'}\over2} \ \cn_a \cn_b \
\int \ d\t \ {\hat \cv}^{\l}(\t )
\ \pr_a \pr_b f_{\l} \ {\bar\z}^a(\t ) {\bar\z}^b(\t ) \ \n_{ab} \ \right )
\nonumber \\
&& \cdot \exp \left (\ -{{i}\over2} \int \ d\t \ d\t'
{j_{\z}}^a(\t') \ {\bar\z}^b(\t ) \ \n_{ab} \right ) 
\nonumber \\
&&
\cdot \exp \left ( \ - {{\a'}\over 2} \int d\t \ d\t' 
\ {\tilde k}_a(\t ) \ G^{-1}_{ab}(\t , \t' ) \
{\tilde k}_b(\t' ) \ \n_{ab} \ \right ) \nonumber \\
&&
\cdot \exp \Big ( \ - {{\a'}\over{3!}} \cn_a^{-1} \cn_l\cn_m\cn_m \ \Big [
K^{\l}_{lm} K_{\l np} \ h^{bp} \ + \ \G^b_{lp} \G^p_{mn} \
+ \ \pr_l\pr_m f^{\l} \pr_q f_{\l} \ \pr_n h^{bq} \nonumber \\
&& 
\qquad\qquad\qquad\qquad + \ 2 \ \pr_l \pr_m f^{\l}
\pr_q f_{\l}\ \G^q_{np} h^{bp} \
- \ \pr_r f^{\l} \pr_q f_{\l} \ \G^q_{np} \G^r_{lm} \ h^{bp} \ \Big ]
\nonumber \\
&& \quad\qquad \cdot \int \ d\t \ d\t' \
{\bar\z}^l(\t ) {\bar\z}^m(\t )
{\bar\z}^n(\t ) \ G_{ab}^{-1}(\t ,\t' )\ {\bar\z}^b(\t' )\ \n_{ab} \ \Big ) \ ,
\label{intzetab}
\eea
where
\be
j_{\z}^a(\t' ) \ \equiv \ {\sqrt{\a'}} \ 
{\tilde k}_b\ G_{ab}^{-1}(\t ,\t' ) \ \d_{ab} \ .
\ee

\sp
To simplify the exponent, which contains a term linear in 
 ${\bar\z}$, let us define $D_{ab}(\t , \t' )$ and ${\tilde\z}^a(\t )$ by 

\bea
D_{ab}(\t , \t' ) \ &=& \ G_{ab}^{-1}(\t , \t' ) 
\ - \ i\a' \ \cn_a \cn_b \ {\hat\cv}^{\l}
(\t ) \ \pr_a \pr_b f_{\l} \comma \\
{\bar\z}^a(\t ) \ &=& \ {\tilde\z}^a(\t ) 
\ - \ i D_{ab}^{-1}(\t , \t' ) \ j_{\z}^b
(\t' ) \ .
\eea
Then the integral becomes 
\bea
I_{\tilde\z} &=& \int \ \cd {\tilde\z}^a(\t ) \  
\exp \left ( \ {1\over2}
\ \int \ d\t \ d\t' \ {\tilde\z}^a(\t ) \ D_{ab}(\t , \t' ) 
\ {\tilde\z}^b(\t' )
\ \n_{ab} \ \right ) \nonumber \\
&&
\quad \cdot \exp \Big ( \ - {{\a'}\over{3!}} \  \cn_a^{-1} \cn_l \cn_m \cn_n \ 
\Big [ K^{\l}_{lm} K_{\l np} \ h^{ap} \ + \ \G^a_{lp} \G^p_{mn} \ + \
\pr_l \pr_m f^{\l} \pr_q f_{\l}
\ \pr_n h^{aq} \nonumber \\
&& \qquad\qquad\qquad\quad + \ 2 \ \pr_l \pr_m f^{\l} \pr_q f_{\l}
\ \G^q_{np} h^{ap} \
- \ \pr_r f^{\l} \pr_q f_{\l} \ \G^q_{np} \G^r_{lm} \ h^{ap} \ \Big ]
\nonumber \\
&&
\qquad\qquad\qquad\quad \cdot \int \ d\t \ d\t' \
{\tilde\z}^l(\t ) {\tilde\z}^m
(\t ) {\tilde\z}^n (\t ) \ G_{ab}^{-1}(\t ,\t' ) 
\ {\tilde\z}^b(\t' ) \ \n_{ab} \  \Big ) \ .
\eea
In the framework of the derivative expansion,  $\co({\tilde\z}^4)$ terms 
 in the exponent  with an extra $\a'$ will be treated perturbatively.
 The basic Gaussian integral gives the determinant 

\bea
\Big ( -\ \det D_{aa} \ \Big )^{-{1\over2}} 
&=& \exp \left ( 
\ -{{i}\over2}\a' \ \cn_a^2 \ \pr_a \pr_a f_{\l} \int d\t \ G_{aa}(\t ,\t )
{\hat\cv }^{\l}(\t ) \ \right ) \ .
\eea
$G_{aa}(\t,\t)$, containing $\ln \epsilon$ divergence, is actually
 independent of 
 $\t$ and the integral fortunately vanishes due to the absence of 
 the zero mode in ${\hat\cv}^{\l}(\t ) $, \ie\   by 
 $\int \, d\t \ {\hat\cv}^{\l}(\t ) \ = \ 0 \ . $ Thus we get  
$\Big ( - \ \det D_{aa} \ \Big )^{-{1\over2}} 
\ = \ 1$.

\sp
Evaluation of  the effect of the quartic interaction is 
 straightforward using  the propagator 
$ \left < \ {\tilde\z}^a(\t ) \ {\tilde\z}^b(\t' ) \ \right >
\ = \ \n^{ab} \ G_{ab}(\t ,\t' ), $ 
which  at the coincident point has the form 
$ 
\left < \ {\tilde\z}^a(\t ) \ {\tilde\z}^b(\t ) \ \right >
\ = \ 2\ \n^{ab} \ h \ \Big ( h - {\bar f}^2 \Big )^{-1} \ln \eps \ . $
After some  algebra, we get a remarkably simple  result:

\be
{\a'} \ln \eps \ h \ \Big ( h-{\bar f}^2 \Big )^{-1}
\cn_a^2 \ K^{\l}_{aa} \ K_{\l ab} \ h^{ab} .
\ee

\sp
Putting everything together, the path integral over the boudary fields
finally yields 
\be
I_{\tilde\z} \ \equiv \ {1\over{\a'}} {\sqrt{-h}} \
\left ( \ 1 \ - \
\a' \ h \Big ( h-{\bar f}^2\Big )^{-1} \cn_a^2\ \n_{aa} \
K^{\l}_{aa} \ K_{\l ab} \ h^{ab} \ \ln \eps \ \right )\ .\label{zint}
\ee
\section{Renormalized vertex operator and its property}
\subsection{Renormalization}

\hs
Substituting the above result (\ref{zint}) into (\ref{bulkint}), 
 we arrive at the disk amplitude to the next to leading order 
 in the derivative expansion:
\be
{1\over{g_s \a'}} {\sqrt{- \Big ( \ h + {\bar F} \ \Big )}}
\ .\left ( \ 1 \ - \ \a' \ \ h \Big ( h-{\bar f}^2\Big )^{-1}
\cn_a^2
\ \n_{aa} \ K^{\l}_{aa} \ K_{\l ab}
h^{ab} \ \ln \eps \ \right ). \label{diskamp}
\ee
The leading term  reproduces precisely the Dirac-Born-Infeld action
 \cite{andreev} for a D-string  with tension $1/(g_s \a' )$.
In our formalism, it is generated from a combination of the 
 bulk integral and the boundary integral. 

The subleading term is the correction due to the boundary interaction. 
It contains the extrinsic curvature of the curved D-string world-sheet 
 and is a non-trivial generalization of the D-particle case. 
In fact one can check that by substituting $\cn_1 = 0$, 
the above amplitude reduces to that
obtained in \cite{hk} for a D-particle.

\sp 
Now we must ask if this divergent correction can be properly absorbed 
 by a renormalization of the D-string coordinate $f(t,\sigma)$, 
 just as in the D-particle case \cite{hk}. The answer is in the affirmative.
Let $f^\mu_R(t,\sigma)$ be the renormalized coordinate which is 
 related to the bare coordinate by 
\be 
f^{\mu} \ = \ f^{\mu}_R  \ + \ \sum_a \d_a f^{\mu}_R \ ,
\ee
where 
\be
\d_a f^{\mu}_R \ = \ -\a' 
\left ( {{ h_R + {\bar F}}\over{h_R}-{\bar F}} \right ) \ \n_{aa}\ \cn_a^2 \ 
K^{\mu}_{aa} \ \ln \eps \ .
\ee
Then, after a long but straightforward calculation, we find, 
 up to ${\cal O}(\a')$, 
\bea
{\sqrt{-  ( h + {\bar F}  )}} &=& {\sqrt{- ( h_R + {\bar F} 
 )}} \left( 1+\a'\, h_R \Big ( h_R-{\bar f}^2\Big )^{-1}
\cn_a^2
\ \n_{aa} \ K^{\l}_{aa} \ K_{\l ab}
h^{ab}_R \ \ln \eps \ \right )
\eea
Putting this into  (\ref{diskamp}), we see that the renormalized 
 expression for the amplitude takes precisely the form of the DBI action 

\be
{1\over{g_s \a'}} {\sqrt{- \Big ( \ h_R \ + \ {\bar F} \ \Big )}} \ .
\label{dbi}
\ee
In the original derivation \cite{leigh}, the action of this form 
 was obtained by quite a different logic. In that treatment, the action 
 was determined so that the equation of motion derived from it 
 gives a flat D-brane, namely the vanishing of the extrinsic curvature. 
In the present treatment, we are dealing with a curved D-string and 
 our result shows that the DBI action is valid including this case if 
 one properly renormalizes its coordinates. 
\subsection{Vertex operator for scattering with tachyons}

\hs
We are now in a position to write down the detailed form of the 
 vertex operator,  defined in (\ref{defvertex}), 
 describing the scattering of tachyons from an arbitrary curved 
 D-string. From  (\ref{bulkint}) and (\ref{dbi}) it takes 
 the form 
\bea
&& \cav_T \Big ( f^{\mu} , \{ k_i\} \Big ) \ = \
{1\over{g_s\a'}} \int dt \ d\s \
{\sqrt{- \Big ( \ h_R + {\bar F} \Big )}}\ 
\cdot \exp \Big ( i k_{\mu} f^{\mu} \Big )
\nonumber \\
&& \qquad\quad
\cdot \int \prod_i ( g_s \ d^2z_i ) \ 
\cdot \exp \left ( {{\a'}\over2} \sum_{ij}{}'  \
k_i^{\mu} k_j^{\nu} \ \left [ h_{\mu\nu} G(z_i , z_j ) \ + P_{\mu\nu} 
D(z_i , z_j ) \right ] \ \right ) \ ,\label{vertex}
\eea
where $k^{\mu} = \sum_i k_i^{\mu}$ is the total momentum of tachyons and
the prime
on the summation implies that the singular part of the Green's function for 
$i=j$ is omitted. 

\sp
In the above expression, there still remain divergences due to 
 the appearance of the bare coordinates 
 $f^{\mu}(t ,\s )$ and its derivatives in the exponent. They are of the 
 following form: The one  in the factor $\exp(ik_\mu f^\mu)$ 
 is proportinal to $k_\mu \delta_a f_R^\mu= k_\mu K^\mu_{aa}$,
 while $h_{\mu\nu}$ in 
 the second exponential contains the piece proportional to 
$ k^{\mu} \pr_a f_{\mu}^R.$ As we shall explain below, they actually vanish 
 due to the \lq\lq current conservation " that follows from the 
 requirement of $SL(2,R)$ invariance of the vertex operator. 

\sp
For simplicity, let us consider the two tachyon case for illustration.
The $SL(2,R)$ transformation can be written as 
\bea 
z &\longrightarrow & \tilde{z}=
{\alpha z + \beta \over {\bar \beta} z + {\bar \alpha}}
\comma \qquad |\alpha^2| -|\beta^2| = 1 
\eea
The measure transforms as $d^2\tilde{z} = d^2z|{\bar \beta} z+{\bar \alpha}
|^{-4}$, while the exponent becomes 
\begin{eqnarray}
&& {\al'\over 2}\Biggl( 2k^\mu_1k^\nu_2 h_{\mu\nu} \Delta G(z_1,z_2) \nn \\
&& + k^\mu_1k^\nu_1 h_{\mu\nu} \Delta G(z_1,z_1) 
+ k^\mu_2k^\nu_2 h_{\mu\nu} \Delta G(z_2,z_2)\nn\\
&& + 2k_1\cdot k_2 D(z_1,z_2) -(k_1^2\ln|1-|z_1|^2| 
 + k_2^2\ln|1-|z_2|^2|)\Biggr) 
\end{eqnarray}
where 
 $\Delta G \equiv  G -D$. 
It is easy to see that the part containing $h_{\mu\nu}$ is invariant 
 if and only if the current conservation condition $k^\mu \del_a f_\mu=0$
 is met. For the rest, $D(z_1,z_2)$ is invariant by itself, while 
 the last two terms cancel the contribution from the measures if and only if 
the tachyon on-shell conditions $k_1^2=k_2^2=4/\al'$ are satisfied.
This result persists for the general multi-particle case and hence 
 the $SL(2,R)$-invariance condition reads 

\bea
&& k_{\mu} \e^{\mu}_a \ \equiv \ k_{\mu} \pr_a f^{\mu} \ = \ 0 \nonumber \\
{\rm and} && \a' \sum_{\a} k_{i\a}^2 \ + \ \a' \sum_a k_{ia}^2 \ -
4 \ = \a' k_i^2 \ - \ 4 \ = 0 \ .
\eea
It is easy to see that with the current conservation condition the 
 divergent pieces vanish and we have a well-defined vertex operator 
 in terms of the renormalized fields. This structure is again quite 
 analogous to the D-particle case discussed in \cite{hk}. 
\section{Attempt at quantized scattering amplitude}

\hs
Having obtained the vertex operator, let us make an attempt to obtain 
 a quantized scattering amplitude. Contrary to the D-particle case, 
 a D-string is an extended object containing infinite number of 
 excitations. Hence the natural amplitude is the one for the 
 scattering of the elementary excitations of D-string with 
 those of F-string. As the effect of the latter is already encoded 
 in the vertex operator itself, what we need to do is to insert 
 the vertex operators for the D-string excitations and then perform the 
 path  integral over the D-string coordinates, \ie\   quantize the 
 D-string. 
\subsection{Brief summary of covariant quantization of D-string}

\hs
Let us begin by making a brief summary of the covariant quantization 
 of D-string, which has been developed over the past 
year\cite{bkop,kal1,kal2,hatkam}. We follow the approach of \cite{hatkam}, which  is most suitable for our purpose. 

\sp 
One begins with the DBI action for a D-string given by 
\begin{eqnarray}
S &=& -\Ttil \int d^2\sigma \sqrt{-h_F} \comma 
\end{eqnarray}
where
\begin{eqnarray}
\Ttil &\equiv & {T \over g_s} \comma \qquad T ={1\over 2\pi \al}\comma \\
h_F &\equiv &\det(h+\bar{F})_{ab} = h+\fbar^2 \comma \\
h_{ab} &=& \del_a f^\mu \del_b f_\mu\comma \qquad \bar{F}_{ab} = \ep_{ab}
\fbar \comma \\
\fbar &=& \ep^{ab}\del_a A_b = \dot{A}_1 -A'_0 \period
\end{eqnarray}
The definitions of the momentum and the electric field are 
\begin{eqnarray}
p_\mu &=& {\del \calL \over \del \dot{f}^\mu} = {\Ttil \over \sqrt{-h_F}}
(\dot{f}_\mu h_{11} -f'_\mu h_{01})\comma \\
E&\equiv & E^1 = {\del \calL \over \del \dot{A}_1}={\Ttil \fbar \over 
 \sqrt{-h_F}}\period 
\end{eqnarray}
From these follow the primary constraints
\begin{eqnarray}
L_\pm &=& \half (p\pm T_E f')^2 = 0 \comma \\
E^0 &=& 0 \comma 
\end{eqnarray}
where $T_E \equiv \sqrt{ \Ttil^2+E^2}$. 
Further, from the consistency of $E^0=0$ with the time-\break
 development, 
 one gets, in the usual manner, the Gauss law constraint 
$ \del_1 E = 0 $.
Under the Poisson bracket, $L_\pm$ can be shown to 
form the left and right Virasoro 
 algebras even in the presence of the electric field. One can then 
 construct the BRS charges and perform the gauge-fixing in the standard 
 manner.  \par
The convenient gauge choice is the conformal gauge defined by 
\begin{eqnarray}
h_{ab} &=& \matrixii{-e^\phitil}{0}{0}{e^\phitil} \comma \label{cfgauge}
\end{eqnarray}
where $\phitil$ is the D-Liouville field.
In this gauge, the action takes the form \cite{hatkam}
\begin{eqnarray}
S &=&\int d^2\sigma \left( p\cdot \dot{f} -{1\over 2T_E}(p^2+T_E^2 f'^2)
 + E \del_0A_1 \right) \period 
\end{eqnarray}
Upon integration over $A_1$, we find that $E$ is a constant.Finally,  further 
integration over the momenta yields the usual Polyakov-type action
\begin{eqnarray}
S &=& {T_E \over 2}\int d^2\sigma \del_a f^\mu \del^a f_\mu \period 
\end{eqnarray}
Since $E$ is constant, so is the D-string tension $T_E$ 
 and hence the quantization of this action can be performed just 
 like for F-string. 
\subsection{Scattering amplitude and the problem of conformal 
 invariance}

\hs
For simplicity, let us take the excitations of the D-string to be 
 tachyons (to be called D-tachyons, to be distinguished from the 
 tachyons of F-string.) Since the action is of the usual Polyakov form, 
 the vertex operator for a D-tachyon carrying momentum $p$ should be, 
 in conformal gauge, 
\begin{eqnarray}
 g_D \int d^2\sigma \sqrt{-h}\, e^{ip\cdot f} &=& 
 g_D \int d^2 \sigma e^{\phitil(\sigma)}  e^{ip\cdot f} \comma 
\end{eqnarray}
where $g_D$ is the D-string coupling constant, which in our present 
 formalism is a free parameter\footnote{It should be determinable in 
 a more complete theory of D-string.} Also,in this gauge $\sqrt{-h_F}$ and 
 $h_{\mu\nu}$ take the following form:

\begin{eqnarray}
\sqrt{-h_F} &=& {\Ttil \over T_E}e^\phitil \comma \qquad 
h_{\mu\nu} =e^{-\phitil} \del_a f_\mu \del^a f_\nu \period 
\end{eqnarray}

The amplitude for 
 the scattering of $N$ tachyons with $M$ D-tachyons then becomes 

\begin{eqnarray}
\calA(k_i, p_I) &=& \int \calD f^\mu \exp\left( -{1\over 4\pi \altilp}
 \int d^2\sigma \del_a f^\mu \del^a f_\mu\right) \nn\\
&& \cdot {T\over g_s }{\Ttil \over T_E} \int d^2u e^{\phitil(u)}
 e^{ik\cdot f(u)}\nn\\
&& \cdot \int \prod_{i=1}^N(g_sd^2z_i) \exp\left( {\al'\over 2}
e^{-\phitil(u)} \sum_{i,j}k^\mu_ik^\nu_j\del_a f_\mu(u)\del^a f_\nu(u)
\Delta G(z_i,z_j)\right)\nn\\
&& \cdot \exp\left( {\al'\over 2}\sum'_{i,j}k_i\cdot k_j D(z_i,z_j)\right)\nn\\
&& \cdot \int \prod_{I=1}^M (g_Dd^2w_I) e^{ip_I\cdot f(w_I)} 
 e^{\phitil(w_I)}  \comma  \label{dfscatamp}
\end{eqnarray}
where, as before, $\Delta G \equiv G-D$. The Neumann function 
 $G(z,z')$, defined in (\ref{diagG}), appears to depend on the D-string 
 coordinate through $h$. However, from the definition of the 
 electric field, which is constant, one easily finds 
$h\pm\fbar^2 = h\left( 1\mp {E^2\over T_E^2}\right)=$ constant and thus 
 $G$ actually does not depend on  $f^\mu(t,\sigma)$. 

\sp
Now we come to face the question of conformal invariance. There are 
 several places where the D-Liouville field $\phitil$ appears and 
 for the theory to be conformally invariant they must be eliminated. 
To study this problem, it is instructive to recall how the Liouville 
 fields disappear in the usual string theory \cite{polyakov}.

\sp
 Consider first the  tachyon vertex 
 insertions. After performing the integral 
 over the string coordinates, one gets the structure 
 \begin{eqnarray}
\prod_i e^{\phi(z_i)} e^{(\al'/2)\sum_{i,j}p_i\cdot p_j \calG(z_i,z_j)}
\comma 
\end{eqnarray}
where  $ \calG(z',z) = \half \ln|z'-z|^2 $ 
is the closed string Green's function.
At the coincident point, this diverges and we must regularize. Define
$
\calG_{reg} (z,z) \equiv  \half \ln |\Delta z|^2 $, where 
 $|\Delta z|$ is the cutoff distance.
What we have to fix actually is the invariant distance 
$ \ep = e^\phi(z)|\Delta z|^2 $.
Thus, 
\begin{eqnarray}
\calG_{reg}(z,z) &=& \half \ln( e^{-\phi(z)}\ep) 
= \half \ln \ep -\half \phi(z) \period 
\end{eqnarray}
Then the dependence on the Liouville field at $z_i$ is 
\begin{eqnarray}
e^{\phi(z_i)}\cdot e^{-(\al'/4)p_i^2 \phi(z_i)} \period 
\end{eqnarray}
Therefore the Liouville dependence disappears  for on-shell tachyon, 
 \ie\ for $p_i^2 = 4/\al'$. (The $\ln\ep$ piece disappears due to the 
 momentum conservation, $\sum_{i,j}p_i\cdot p_j = 
 (\sum_ip_i)^2 =0$.)\par

\sp
Consider next the case of the graviton-dilaton vertex. The vertex is 
 of the form 
\begin{eqnarray}
\int d^2z\sqrt{-h}\, \ep^{\mu\nu}\del_a X_\mu \del_b X_\nu h^{ab}e^{ik\cdot X}
&=& \int d^2z e^{\phi}\left\{ e^{-\phi} \ep^{\mu\nu}\del_a X_\mu \del^a X_\nu
e^{ik\cdot X}\right\}
\nn\\
&=& \int d^2z\ep^{\mu\nu}\del_a X_\mu \del^a X_\nu e^{ik\cdot X}\period 
\end{eqnarray}
So it does not have the Liouville factor. This in turn means that the 
Liouville dependence from the Green's function must vanish by itself.
This requires precisely the on-shell condition for the graviton-dilaton,
\ie\ $k^2=0$. Similar mechanism works for higher rank tensor fields
\footnote{Such a mechanism has been implicitly applied when we performed the 
 integration over the F-string. Omission of the contribution from
 the singular part 
 of the Green's function at the coincident point  is the consequence of 
 this mechanism.}. 

\sp
Now let us go back to our amplitude (\ref{dfscatamp}). If we expand 
 the exponential in the third line and combine 
 each of the term with the factor $\int d^2u e^{\phitil(u)}
 e^{ik\cdot f}$ in the second line,  we see that our vertex operator 
 describing the interaction of D- and F- strings consists of an 
 infinite sum of vertex operators for tensor fields of various ranks. 
They come precisely with the expected D-Liouville factors discussed 
 above for the ordinary string. 
Each vertex operator is thus conformally invariant for the respective on-shell 
 value of $k^2$, namely $k^2 = -4n/\altilp\comma \ n=-1,0,1,\ldots$, 
 where $\altilp = 1/2\pi T_E$ is the slope parameter for D-string. 
This means that our vertex operator is dominated by the effect of 
 an infinite collection of on-shell resonances. This unfortunately does not 
 respect the conformal invariance as a whole. 
\par
This must have occured due to our approximation. In our scheme, we have 
 started with the contribution of the configuration where the ends of the open 
 string are attached to a point on the D-string worldsheet and then 
 tried to take into account the fluctuation from this limit in the 
 derivative expansion. The first correction so obtained is  
 the factor $\sqrt{-(h+\bar{F})}$ (and its renormalization).
 Therefore, up to this level, the interaction is essentially point-like 
 and hence only the effects of the on-shell intermediate states are picked up.
\par
 Thus we have learned that , unlike in the case of a D-particle 
interacting with 
 a string, the requirement of symmetry 
 governing the system of two different types of extended objects is 
 much more stringent and is non-trivial to implement. Our work, which 
attempted to go beyond the existing knowledge by quantizing the D-string 
 coordinates, revealed this feature in an explicit manner. 
\par
 How would one overcome this difficulty ? There may be several 
 directions for progress:
\begin{itemize}
\item One way is to try to investigate the 
 effect of  higher order corrections in the present scheme: One should then be 
 able to see how the off-shell intermediate states would begin to 
 contribute and  may get a hint for constructing satisfactory 
 off-shell vertex operator. 
\item  Alternatively, one may look for a scheme  which 
 would automatically guarantee the conformal invariance. Imposition of 
 BRST invariance in the operator approach may be among such possibilities. 
 In this piture, the amplitude that we attempted to compute is represented 
 by the expression of the form 

$${}_D\langle 0 | V_D(p_1) \cdots 
 V_D(p_M)\ {\cal V}\  V_F(k_1) \cdots V_F(k_N)|0\rangle_F\comma  $$
where ${}_D\langle 0 |$ and $V_D(p_i)$ are, respectively, the vacuum state 
 and the vertex operator in the D-string sector and similarly 
 the ones with the subscript $F$ denote the corresponding quantities 
 in the F-string sector. In the middle is the vertex operator $\cal V$ 
 which converts between D- and F- sectors, just like the fermion emission 
 vertex of the RNS formalism of superstring theory. For conformal (or BRST) 
 invariance, $\cal V$ must satisfy $Q_D {\cal V} = {\cal V} Q_F$, where 
 $Q_D$ and $Q_F$ are, respectively, the BRST charge in the D- and F- sector.
 This equation should restrict the form of $\cal V$ considerably and 
 may even  be powerful enough to determine  $\cal V$. 
\end{itemize}

The studies suggested above  are, however,  beyond the scope of the present 
investigation and are left for the future. 

\vspace{.3in}
\noindent
{\large\bf Acknowledgments } \par
S.K. is grateful to the members of the High Energy Physics Theory
Group in the Institute of Physics for various help. His work is supported 
 by the grant from the Japan Society for the Promotion of Science, JSPS-P96012.
Y.K. would like to thank T.Yoneya for a useful discussion. He is also 
 grateful to the organizers and the participants of the duality workshop
 held at ITP, Santa Barbara for hospitality, 
where a part of this work was performed.  
The work of Y.K. is  supported in part by the Grant-in-Aid for 
Scientific Research (No.09640337) from the Ministry of Education, 
Science and Culture and by the National Science 
Foundation under Grant No.PHY94-07194.

\def\anp#1{Ann. of Phys.}
\def\prl#1{Phys. Rev. Lett.}
\def\prd#1{{Phys. Rev.} {\bf D#1}}
\def\plb#1{{Phys. Lett.} {\bf B#1}}
\def\npb#1{{Nucl. Phys.} {\bf B#1}}
\def\mpl#1{{Mod. Phys. Lett} {\bf A#1}}
\def\ijmpa#1{{Int. J. Mod. Phys.} {\bf A#1}}


\end{document}